%% file: emapj.tex
\documentclass{emulateapj}

\newcommand{\htwo}{H$_2$}
\newcommand{\ebv}{$E(B-V)$}

\shortauthors{BURGH, FRANCE \& McCANDLISS}
\shorttitle{CO/H$_2$ IN THE DIFFUSE ISM}

\begin{document}

\title{Direct Measurement of the Ratio of Carbon Monoxide to Molecular Hydrogen in the Diffuse Interstellar Medium}

\author{Eric B. Burgh}
\affil{Space Astronomy Laboratory, University of Wisconsin - Madison\\
1150 University Avenue, Madison, WI 53706}
\email{ebb@sal.wisc.edu}
\author{Kevin France\altaffilmark{1}, Stephan R. McCandliss}
\affil{Department of Physics and Astronomy, The Johns Hopkins University\\
3400 North Charles Street, Baltimore, MD 21218}

\altaffiltext{1}{Current Address: Canadian Institute for Theoretical Astrophysics, University of Toronto, 60 St. George Street, Toronto, Ontario, M5S 3H8}


\journalinfo{Accepted for publication in The Astrophysical Journal}
\submitted{}

\begin{abstract} 
We have used archival far-ultraviolet spectra from observations made
by the Space Telescope Imaging Spectrograph (STIS) of the
\textit{Hubble Space Telescope} and the \textit{Far Ultraviolet
Spectroscopic Explorer} (\textit{FUSE}) to determine the column
densities and rotational excitation temperatures for carbon monoxide
and molecular hydrogen, respectively, along the lines of sight to 23
Galactic O and B stars.  The sightlines have reddening values in the
range \ebv~=~0.07--0.62, thus sampling the diffuse to translucent
interstellar medium.  We find that the \htwo\ column densities range
from $5\times10^{18}-8\times10^{20}$ cm$^{-2}$ and the CO from upper
limits around $2\times10^{12}$ cm$^{-2}$ to detections as high as
$1.4\times10^{16}$~cm$^{-2}$.  CO increases with increasing \htwo,
roughly following a power law of factor $\sim2$. The CO/\htwo\ column
density ratio is thus not constant, and ranges from $10^{-7}-10^{-5}$,
with a mean value of $3\times10^{-6}$.  The sample segregates into
"diffuse" and "translucent" regimes, the former having a molecular
fraction less than $\sim0.25$ and $A_V/d<1$~mag~kpc$^{-1}$.  The mean
CO/\htwo\ for these two regimes are $3.6\times10^{-7}$ and
$9.3\times10^{-6}$, respectively.  These values are significantly
lower than the canonical dark cloud value of $10^{-4}$.  In six of the
sightlines, the isotopic variant $^{13}$CO is observed, and the
isotopic ratio we observe ($\sim50-70$) is consistent with, if perhaps
a little below, the average $^{12}$C/$^{13}$C for the interstellar
medium at large. The average \htwo\ rotational excitation temperature
is $74\pm24$ K, in good agreement with previous studies, and the
average CO temperature is 4.1 K, with some sightlines showing
temperatures as high as 6.4 K.  The higher excitation CO is observed
with higher column densities, consistent with the effects of photon
trapping in clouds with densities in the 20-100~cm$^{-3}$ range. We
discuss the implications for the structure of the diffuse/translucent
regimes of the interstellar medium and the estimation of molecular
mass in galaxies.
\end{abstract}


\received{26 July 2006}
\revised{9 November 2006}
\accepted{21 November 2006}

\section{Introduction}

Molecular hydrogen (\htwo) is the most abundant molecule in the
interstellar medium (ISM), residing primarily in the large complexes
of the dense molecular clouds that account for 10-20\% of the mass in
the inner disk of the Galaxy \citep{Shull82}.  However, it is
difficult to observe directly.  \htwo\ is a homonuclear molecule, with
quadrupolar ground-state transitions that emit radiation only very
weakly.  Carbon monoxide (CO), in contrast, has strong ground-state
transitions that produce readily observable emissions at radio
wavelengths.  In clouds with densities above the critical density for
CO, \htwo\ collisions dominate the ground-state excitation, and thus
CO radio emission acts as a tracer of \htwo\ in these regions.

\begin{figure*}
\plotone{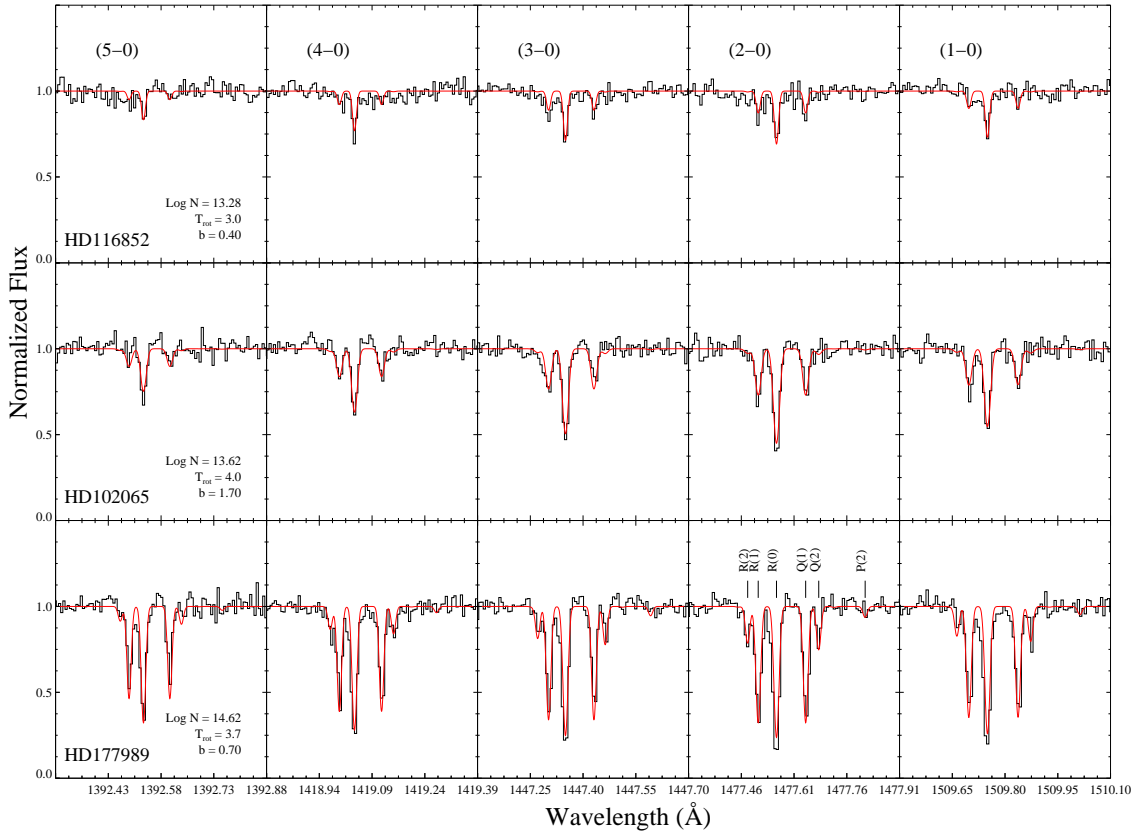}
\caption{Sample CO absorption profiles for sightlines with increasing column
density.  The best fitted values for column density, rotational excitation temperature and Doppler broadening parameter are listed.
\label{cosample}}
\end{figure*}

This relationship is often characterized by the conversion factor $X =
N(\mathrm{H_2})/I_\mathrm{CO}$, where $I_\mathrm{CO}$ is the
integrated brightness temperature of the $J=1-0$ radio
emission line at 2.6 mm.  The value for the conversion factor is
generally determined by one of the following techniques
\citep[cf.][]{Young91}: correlation of the CO emission with $A_V$ in
clouds determined by star counts, which is then correlated with \htwo\
by the extrapolation of the $N_H/A_V$ from the diffuse ISM
\citep{Savage77}; an excitation analysis of $^{13}$CO, assuming it is
optically thin, $^{12}$CO is optically thick, and the
$^{12}$CO/$^{13}$CO is known; a virial analysis using the cloud sizes
and linewidths; and comparison with $\gamma$-ray emission
\citep[e.g.][]{Strong96}.  These techniques produce a value for $X$ of
about $2\times10^{20}$~cm$^{-2}$~(K~km~s$^{-1}$)$^{-1}$, but examples
of up to a factor of 10 higher can be found in the literature.

It is generally recognized that CO emission can qualitatively trace
the distribution of \htwo\ in the ISM, but because of difficulties in
translating a measurement of $I_\mathrm{CO}(1-0)$ into a column
density of CO, $X$ is not a good quantitative measure of
$N(\mathrm{CO})/N($\htwo$)$ (hereafter CO/\htwo).  To relate $X$ to a
relative column density ratio requires the assumption of LTE and is
typically done with the unsaturated $^{13}$CO radio lines, so one must
also assume excitation temperatures and abundance ratios for the two
CO isotopes.  This exercise gives a value of CO/\htwo\ of about
$10^{-4}$ for the dense molecular clouds, but with a large uncertainty
\citep[e.g.][]{Dickman78}.

The determination of CO/\htwo\ is simplified by the direct observation
of the two species in absorption.  This has been done in the infrared,
using vibrational transitions, and values of $\sim2-3\times10^{-4}$
were found for the molecular clouds in NGC~2024 and NGC~2264
\citep{Lacy94}.  With ultraviolet observations, this direct method can
be extended to the diffuse and translucent phases of the ISM, where
the molecules are more susceptible to the effects of
photodissociation, and thus CO/\htwo\ is expected to vary relative to
the dense clouds.  Models suggest that the photodissociation of CO is
sensitive, even more so than \htwo, to the strength of the
interstellar ultraviolet radiation field, cloud geometry and the
ultraviolet absorption and scattering properties of dust
\citep{vanDishoeck88,Kopp00}.  Therefore the inter-relationship
between CO and \htwo\ may also be a good measure of the physical
conditions and structure of the diffuse and translucent regimes of the
ISM.  Furthermore, there is evidence that a significant contribution
to the large-scale Galactic CO emission is made by lower optical depth
gas \citep{Polk88}.
 
\htwo\ is best observed in the far-ultraviolet where there are
extensive dipole-allowed electronic absorption band systems.  The
\textit{Far Ultraviolet Spectroscopic Explorer} (\textit{FUSE}) has
proven to be an excellent instrument for studying \htwo\ in absorption
\citep[and others]{Rachford02,Tumlinson02,Gillmon06}.  CO also has an
absorption band system throughout the ultraviolet, and a study of CO
in the diffuse ISM using \textit{Copernicus} was performed by
\citet{Federman80} and later reanalyzed by \citet{Crenny04}.  These
observations were limited to the C-X (1088 \AA) and E-X (1076 \AA)
absorption bands toward 48 nearby bright stars with only about one
third having CO, some of those being saturated.  Further, the spectral
resolution ($R\sim20000$) was not high enough to resolve the
rotational substructure of the bands.  \citet{Federman80} concluded
that measurements with a more sensitive, higher spectral-resolution
instrument would allow for a better understanding of the relationship
between \htwo\ and CO across a wider range of environments.

This can be achieved using the Space Telescope Imaging Spectrograph
(STIS) on board the \textit{Hubble Space Telescope}, which has access
to the Fourth Positive (A$^1\Pi$ -- X$^1\Sigma^+$ $v''-0$) band
system.  More than a dozen absorption bands from the ground
vibrational state of CO have been detected and, because of the large
variance in oscillator strengths, they probe a variety of optical
depths, allowing for an accurate determination of column density.
Additionally, with the highest resolution grating modes (E140H) these
bands can be resolved clearly into their constituent rotational
structure.  Because the rotational levels in the ground state of CO
are closely spaced, the rotational excitation temperature of the
molecule is easily determined.

\citet{Pan05} used STIS Echelle and \textit{FUSE} data to explore the
CO and \htwo\ absorptions in the Cepheus OB2 and OB3 clusters.  They
find evidence for systematic variations in CO/\htwo\ in these two
different star-forming regions, which they say may indicate
differences in star-formation histories.  In this paper, we broaden
the study of CO/\htwo\ to the diffuse molecular regime of the ISM,
rather than study isolated regions.  We present an analysis of 23
stars, which have been observed by both \textit{FUSE} and the STIS
E140H mode.  The reddenings of these stars range from
\ebv$=0.07-0.62$, complementing the Pan et al. study well, whose
sightlines range from \ebv$=0.35-0.86$.  The \htwo\ and CO absorptions
have been measured and column densities and rotational temperatures
determined. These data provide a survey of the CO/\htwo\ relationship
in the diffuse to translucent molecular regime of the ISM.

\section{Data and Analysis}
The data presented here were retrieved from the Multimission Archive
at STScI (MAST).  The STIS observations employed the E140H grating,
providing the highest possible spectral resolution.  The wavelength
coverage of STIS allows for the observation of a number of absorption
bands of the CO A-X ($v''-0$) band system.  For some of the stars,
though, the specific tilt of the Echelle grating allows for
observation of only higher vibrational bands ($v''\geq7$), because of
the more blueward wavelength coverage.  The \textit{FUSE} data were processed 
with the CalFUSE pipeline, version 2.2.  The individual channels (e.g. LiF 
1a, SiC 1a, etc.) were joined using an IDL shifting routine written to
combine time-tagged or histogrammed \textit{FUSE} data.  Additionally, we
obtained STIS data corresponding to five sightlines from a \textit{FUSE} 
survey of translucent clouds \citep{Rachford02}, from whom we obtained the
\htwo\ column densities.

\subsection{Carbon Monoxide}

CO has an extensive electronic absorption band system, the Fourth
Positive ($A^1\Pi-X^1\Sigma^+$) system, ranging from 1510 \AA\ to
shorter wavelengths.  Because the energies of the rotational levels in
the ground state of CO are closely spaced, the relative strengths of
the individual ro-vibrational transitions are very sensitive to
excitation temperature.  Previous surveys, such as Federman et
al. (1980) used \textit{Copernicus} data, which were not able to
resolve the rotational structure but the high spectral resolution of
the STIS E140H grating ($R=\lambda/\Delta\lambda=110,000$) allows
these transitions to be well-resolved from each other and provides for
an accurate determination of the rotational temperature in addition to
column density.

Model spectra of the ro-vibrational absorption were generated using
the wavelengths and oscillator strengths of \citep{Morton94}.  The
best fit model was chosen using a $\chi^2$ statistic from a grid
corresponding to varying column density ($N$), rotational excitation
temperature ($T_\mathrm{rot}$), and Doppler line broadening parameter,
$b$.  For most of the low-column sightlines, the bands (1-0) through (5-0)
were fit simultaneously.  For sightlines with only the
shorter wavelength data, the bands starting with (7-0) and higher were fit.
Figure \ref{cosample} shows a sample of spectra with five bands of the
A-X system range in column densities with corresponding best fit
models overlaid.  When the lower vibrational bands were available, the
detection limit was typically $N\sim2.5\times10^{12}$ cm$^{-2}$.  For
the (7-0) band, it was $N\sim4.0\times10^{13}$ cm$^{-2}$.  In all but
four of the sightlines, CO absorption was detected, and in six of the
higher column sightlines the isotopic variant $^{13}$CO was also
observed.

\begin{figure}[t]
\epsscale{1.1}
\plotone{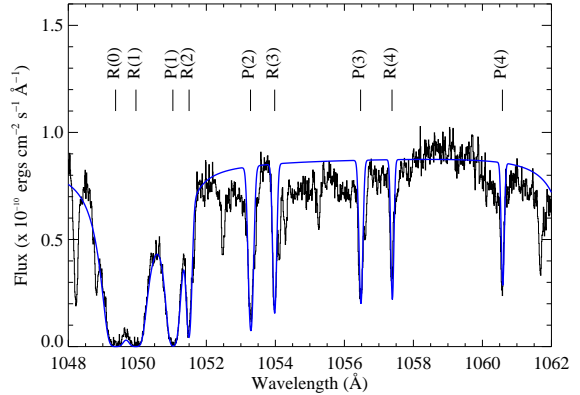}
\caption{Sample \htwo\ absorption (HD~93205) of the Lyman (4-0) band
system, with best fitted absorption profile overplotted.  Absorptions
out of the $J$~=~0~--~4 are included in the fit; however, the $J$~=~0
and 1 lines contain the majority of the column density, and only these
are considered in determining the total \htwo\ column density.  Other
lines in the spectrum are from \ion{Ar}{1} (1048.2\AA), \ion{Fe}{2}
(1055.3\AA), and high J states of the Lyman (5-0) band.
\label{hd93205h2}}
\end{figure}

In addition to the profile fitting, a curve-of-growth (COG) analysis
was performed for each sightline.  The individual ro-vibrational
absorption profiles of each available band were fitted with Gaussians,
their equivalent widths measured, and a COG constructed.  A similar
grid-search process as used in the profile fitting was employed and
the $N$, $T_\mathrm{rot}$, and $b$ determined by comparison to the
theoretical COG for a single velocity component.  For sightlines with
column densities below 10$^{14}$~cm$^{-2}$, the best fitted values
agree with those determined from the profile fitting process within
the error.

\begin{figure*}
\plotone{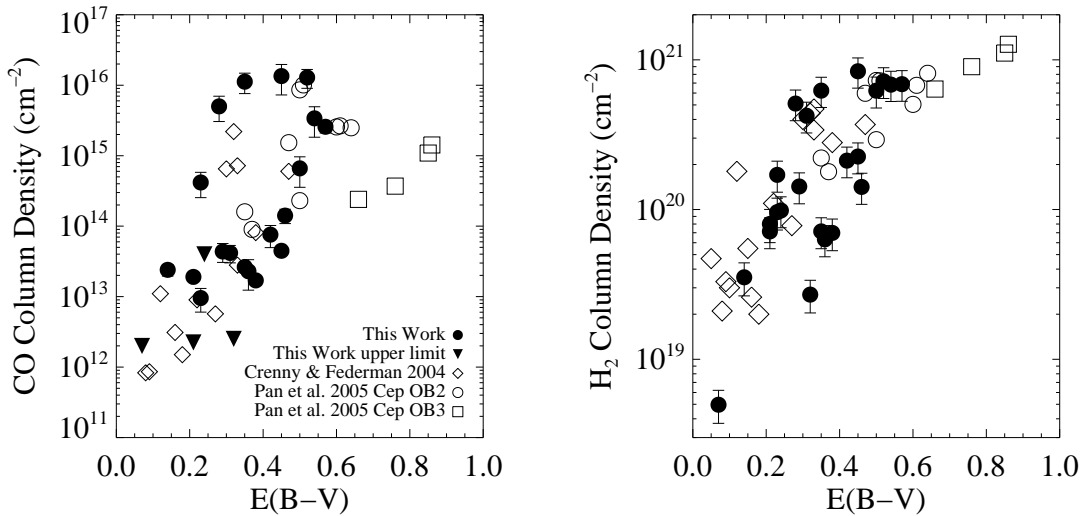}
\caption{CO (left) and \htwo\ (right) column density versus reddening,
\ebv.
\label{plotvebv}}
\end{figure*}

For the higher column sightlines, unresolved velocity components may
influence the results.  The equivalent width of an absorption line can
be higher for a given total column density if the velocity structure
contains unresolved non-overlapping absorption lines.  This effect is
more pronounced as the lines become saturated, and thus the
equivalent widths of the higher oscillator strength bands are
inflated, producing a COG that mimics that of a single component with
higher $b$.  However, if enough unsaturated lines can be observed,
then the column density determination is still robust.  The
combination of the STIS signal-to-noise ratio and high spectral
resolution allow, in virtually all cases, for the observation of
absorption lines that lie on the linear part of the COG.  We noted
that, for values of $b$ above $\sim0.6$~km~s$^{-1}$, saturation
effects are mild to negligible  for absorption lines with an
$Nf\lambda\lesssim10^{15}$~cm$^{-2}$~\AA.

Our initial fits of the higher column sightlines produced $b$-values
in the range $1.0-3.0$.  These are higher than expected for molecular
material (\citet{Pan04} find $b=0.6-1.0$ for CN, for example).  We
found that by limiting the fitting process to only the weaker
absorptions, we were able to reproduce the lower $b$ values.  In these
cases, the fit was limited to those lines with
$W_\lambda/\lambda\sim5-9\times10^{-6}$, depending on the data
quality, to avoid saturation effects. The application of the
equivalent width cut lowered the determined $b$ value, and increased
the determined $N$ by typically 0.2 dex.  The need for this approach
was demonstrated best by our attempts to fit the HD~24534 ($\chi$ Per)
sightline, the $N(CO)$ of which was increased by 0.5 dex after the
application of the equivalent width cut.  Twelve bands of the A-X
system were clearly observed, but we could get no consistent fit from
profile fitting.  Also, the COG shows deviations from what would be
expected for a single set of fit parameters.  This is most likely
because of the presence of unresolved velocity components.

The CO absorption profiles for $\chi$ Per have been fit before using
GHRS data of the A-X system \citep{Kaczmarczyk00} as well as STIS data
of the intersystem bands \citep{Sheffer02}.  \citet{Kaczmarczyk00}
found that a two-component model, with one component having 85\%\ of
the column, was necessary to get a good profile fit of the data.  He
derives a column of $N(\mathrm{CO}) =
(1.0\pm0.2)\times10^{16}$~cm$^{-2}$, consistent with that found by
\citeauthor{Sheffer02} ($1.41\times10^{16}$~cm$^{-2}$) as well as this
study ($1.35\times10^{16}$~cm$^{-2}$).  In principle, the unsaturated
intersystem bands, as \citet{Sheffer02} used, could also be used for
the highest column density sightlines, but there were no STIS E140H
data available in the archive with the appropriate wavelength coverage
for any of the other stars.

\subsection{Molecular Hydrogen}

\htwo\ has an extensive band system in the \textit{FUSE} wavelength range
(912~$\lesssim$~$\lambda$~$\lesssim$~1110 \AA\ at temperatures typical
of the diffuse ISM) arising from electronic transitions out of the
ground state ($X^{1}\Sigma^{+}_{g}$) to the Lyman
($B^{1}\Sigma^{+}_{u}$) and Werner ($C^{1}\Pi_{u}$) levels.  These
absorption lines were first used by \textit{Copernicus} to characterize the
average properties of the molecular phase of the ISM
\citep{Spitzer74}.  More recently, \textit{FUSE} has been used to compile
large samples of \htwo\ absorption data for galactic
\citep{Rachford02,Gillmon06} and extragalactic \citep{Tumlinson02}
sightlines.

We have analyzed the \htwo\ absorption lines in the sightlines
presented here with a least-squares fitting routine that compares the
data to a model spectrum.  The model spectrum was constructed with a
power-law fit to the stellar continuum over the wavelength region of
interest ($\approx$~1035~--~1140~\AA) with absorption components
derived using the $H_{2}ools$ optical depth templates of
\citet{McCandliss03}.  The continuum was placed interactively as
individual adjustments were necessary in almost all cases.  The
optical depth templates are calculated for each line using a Voigt
profile, ensuring the correct line shape for different column density
regimes.  Figure \ref{hd93205h2} shows a sample absorption spectrum of
\htwo\ and the best fitted model.

An iterative approach is used for absorption lines from rotational
states from $J$~=~0 to $J$~=~4.  A coarse grid of column densities
(15.6~$<$~log $N$(H$_{2}$($J$))~$<$~23.0) is scanned for $b$-values
ranging from 2 to 11 km~s$^{-1}$.  This process establishes the column
to within 0.25 dex and $b$-value to within $\pm$~1~km~s$^{-1}$.  A
fine grid of column densities ($\Delta N=0.01$ dex) is then scanned
and the best fit is quoted to the nearest 0.05 dex.  We feel that this
is a conservative estimate of the accuracy of our H$_{2}$ column
density determination.  As a consistency check, we compare our results
for two sightlines also studied by \citet{Rachford02}.  The column
densities measured for HD 185418 are identical to theirs within the
errors.  On the other hand, HD 102065 is discrepant in log $N$(0) by
0.2 dex.

This iterative approach proved repeatable for determining column
densities for the $J$~=~0 and 1 lines.  The higher-lying rotational
states ($J$~=~2~--~4) were fit with the same procedure, initially
scanning a grid of column densities from $10^{13}-10^{20}$
cm$^{-2}$.  The quantitative results for these lines were less certain.
Most of the lines arising from $J$~=~3 and 4 fell on the ``flat'' part
of the curve of growth (as well as those from $J$~=~2 in a few cases),
and were subject to the degeneracy between column density and
$b$-value.  Additionally, some of the lines from $J$~=~2 and 3 were
blended, and we found that continuum placement was even more
significant when dealing with these lines.  To avoid these
complications, we only present the columns derived for the $J$~=~0 and
1 levels as these states contain the majority of the molecular
mass at diffuse ISM temperatures.

\section{Results}

\subsection{Column Densities and CO/\htwo}

Table 1 summarizes the derived column densities and rotational
excitation temperatures for the CO and \htwo\ along the line of sight
to each star, as well as the spectral type and reddening.  For all but
three of the sightlines in this study, we have obtained from the
literature values for the neutral hydrogen column density; the
\ion{H}{1} column densities for HD~27778, HD~102065 and HD~203532 were
determined from Ly$\alpha$ fits to the STIS data.  The spectral type
for HD~102065 given in the literature of B9IV is inconsistent with the
presence of weak but notable \ion{C}{4} $\lambda\lambda$1548--1550
features typical of spectral types B1.5V or slightly later for high
luminosity class \citep{Walborn95}.  Consequently, we adopt a B2V
spectral type for this object ($(B-V)_o=-0.24$), yielding an
\ebv~=~0.31.  Table 2 lists the column densities of the first two
rotational states of \htwo, the column density of \ion{H}{1} and the
molecular fraction, $f=2N($\htwo$)/(2N($\htwo$)+N($\ion{H}{1})).

With both molecules there is a trend toward higher column with
increased reddening (see Figure \ref{plotvebv}), though neither
exhibits a tight correlation; for any given \ebv, the column density
of both molecules can vary by at least an order of magnitude.  The
scatter is most likely a product of varying environments.  The color
excess is a measure of the total amount of dust along the line of
sight, but may not necessarily reflect the geometrical distribution of
the gas and dust; a single translucent cloud could produce the same
reddening as a series of diffuse clouds -- see, for example, the models of
\citet{Kopp00}.

\begin{figure}[t]
\epsscale{1.2}
\plotone{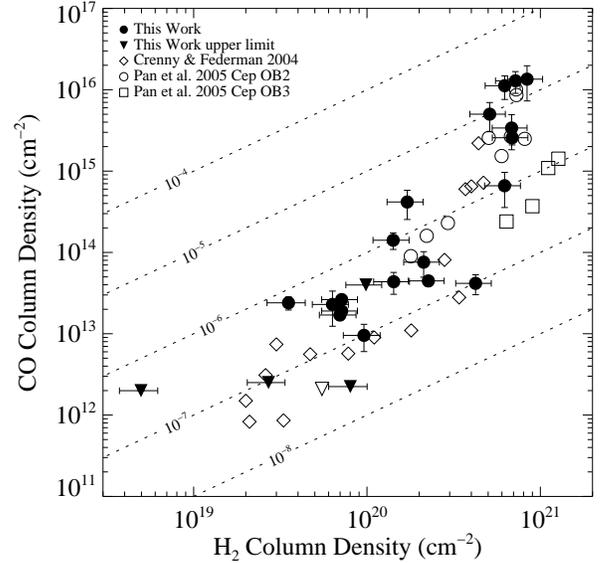}
\caption{Correlation plot of CO with \htwo\ for the data presented in
this study as well as previous studies.  Contours of constant
CO/\htwo\ are overplotted.
\label{ploth2co}}
\end{figure}

Figure \ref{ploth2co} shows the correlation of CO with \htwo.  This is
the tightest correlation seen in this study, roughly a power law
relationship, i.e. $N(\mathrm{CO})\propto N(\mathrm{H}_2)^\alpha$, with
$\alpha\approx2$.  At lower \htwo\ columns, the CO can be difficult to
observe, whereas along the higher column sightlines, CO increases very
quickly.  This is most likely an indication of the onset of CO
self-shielding, but may also be attributed to increased shielding of
CO from photodissociation by either \htwo\ or dust.  Overplotted are
the results from the survey by \citet{Crenny04}, which used
\textit{Copernicus} data and the Cep OB data from \citet{Pan05}, who
used STIS and \textit{FUSE} data similarly to this work.  Pan et
al. note a systematic offset of the CO/\htwo\ relationship between the
Cep OB2 and OB3 clusters, further indicating that differences in the
physical conditions of the clouds are playing a strong role in the
scatter in this relationship.

Contours show lines of constant CO/\htwo.  The slope of the
correlation is steeper than 1 and thus there is an increasing trend of
the CO/\htwo\ with increasing column, with a steepening of the
slope near a CO column of $\approx10^{15}$~cm$^{-2}$, where the self-shielding
of CO becomes important.  The value of CO/\htwo\ reaches
about $10^{-5}$ for the highest column sightlines, and if the general
trend is extrapolated to high \htwo\ column agrees well with the value
for dense clouds of about $3\times10^{-4}$ as determined by
\citet{Lacy94} for NGC 2024 and NGC 2264.  This trend is well
supported theoretically, such as by the models of
\citet{vanDishoeck88}, which suggest that the column density ratio
should increase from around $10^{-7}$ in diffuse clouds to $10^{-4}$
in dense clouds.

\begin{figure*}
\plotone{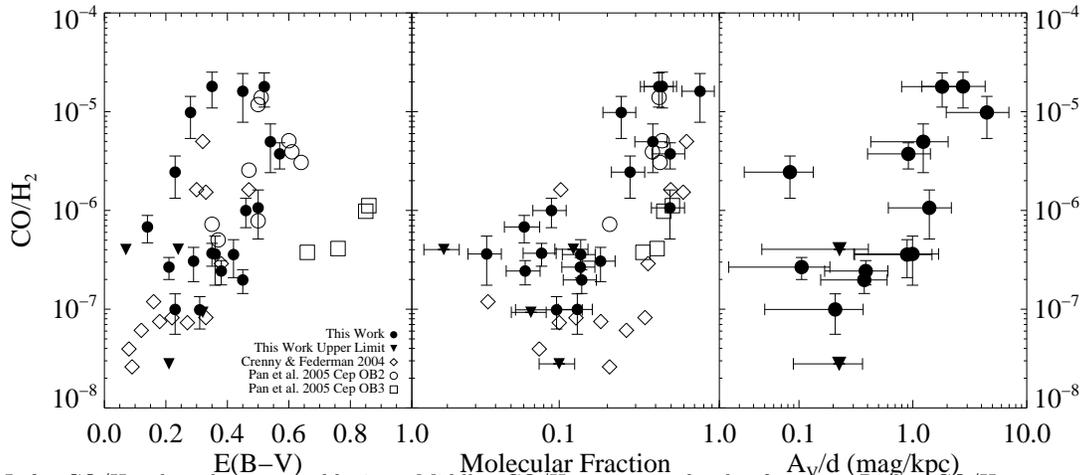}
\caption{Left: CO/\htwo\ plotted versus reddening.  Middle: CO/\htwo\ versus molecular fraction. Right: CO/\htwo\ versus average density ($A_V/d$).
\label{h2coratio}}
\end{figure*}

The left panel of Figure \ref{h2coratio} shows CO/\htwo\ as
a function of reddening.  Again the results of earlier studies are
shown overplotted.  There is a general trend toward
higher ratio with increased reddening.  An interesting standout from
this trend are the Cep OB3 data, which appear to have far smaller
CO/\htwo\ for the \ebv\ than expected.

The middle panel of Figure \ref{h2coratio} shows the ratio with respect
to the molecular fraction.  Below a molecular fraction of about
0.25 (the average value found in the \citealt{Savage77} study), the
CO/\htwo\ is about $3\times10^{-7}$ on average; we will consider
this the ``diffuse'' regime.  These lines of sight have a significant
spread in reddening, covering $0.07\leq E(B-V)\leq 0.46$, and are not
well differentiated from the high CO/\htwo\ sightlines in the
plot versus reddening.  The fact that they separate well from the high
ratio sightlines in this plot suggests that the molecular fraction
represents a better measure of a cloud's physical condition than
\ebv. This is most likely because the amount of dust tends to follow the
total hydrogen along the line of sight rather than the molecular
component and the destruction of these molecules is more sensitive to
line-shielding than dust-shielding.  Above a molecular fraction of
0.25 there is marked increase in the ratio with increasing molecular
fraction, with an average value of about $7\times10^{-6}$.

We see a similar behavior of the CO/\htwo\ with the traditional
density measure $A_V/d$, where $A_V$ is the magnitudes of extinction
in the $V$ band and $d$ is the distance to the star.
\citet{Valencic04} have compiled a large catalog of extinction
properties to reddened Galactic O and B stars, and there is an overlap
of 15 stars with this study.  The right panel of Figure
\ref{h2coratio} shows CO/\htwo\ versus this density
measure. \citet{Jenniskens93} found that $A_V/d = 0.9$~mag~kpc$^{-1}$
appears to separate the diffuse from the dense lines of sight, and our
data show an increase in the CO/\htwo\ above $A_V/d\sim1$; however, we
would state that the division is between diffuse and translucent, as
the sightlines in our study with $A_V/d>1$~mag~kpc$^{-1}$ still have
CO/\htwo\ about an order of magnitude less than that for dense clouds.

Further comparison to the \citeauthor{Valencic04} catalog produces no
significant correlations of extinction curve parameters to the
measured values of $N$ or $T_\mathrm{rot}$ for either CO or \htwo, or
for CO/\htwo.  The study of \citet{Burgh00} indicated a correlation
between the CO column density normalized by \ebv\ and the strength of
the far-UV rise; however, that study, using data from the
\textit{International Ultraviolet Explorer} satellite, covered column
densities ranging from $N$(CO)=$10^{14}-10^{18}$~cm$^{-2}$, the
correlation only becoming apparent at the highest columns and steepest
UV curves.  It is likely that in the environments of the clouds
probed in this study, shielding by \htwo\ and self-shielding are more
important in regulating the abundance of CO than dust-shielding, which
may not play a significant role until $N$(CO)$>10^{16}$.

\subsection{Isotopic Fractionation}

For six of the sightlines, absorption lines from the isotopic variant
$^{13}$CO are observed.  Table 3 lists the $^{12}$CO and $^{13}$CO
column densities and rotational temperatures for these sightlines.
The first column of $^{13}$CO column densities are those determined if
it is assumed that the two isotopes have the same rotational
temperature.  Also listed are the results if the temperature is
allowed to be a free parameter.  The results give column densities
within 0.2 dex of each other.  The next section will discuss the
observed difference in rotational temperatures, when that parameter is
left free to be fit.

The isotopic ratio $^{12}$CO/$^{13}$CO ranges from about 50-70, with
an average value of $57\pm7$ and we see no strong correlations with
any other measurable quantity.  The largest value is for $\chi$~Per
(HD~24534), whose ratio was determined by \citet{Sheffer02} to be
$73\pm12$, in good agreement with the value found here; however, we
do not see the enhanced fractionation measured along such lines of
sight as those to $\rho$~Oph~A, $\chi$~Oph \citep{Federman03}, and
$\zeta$~Oph \citep{Lambert94}, with values of $125\pm23$, $117\pm35$,
and $\sim170$, respectively.  

\citet{Wilson94} review the literature for determinations of the
interstellar $^{12}$C/$^{13}$C and adopt a value of $77\pm7$; however
there is significant scatter amongst the values reported in the
literature that seems to depend on the methods used, which can include
from millimeter emissions of C$^{18}$O (e.g. \citet{Langer93} get
$^{12}$C/$^{13}$C$=57-74$) and near-infrared spectroscopy of CO
vibrational bands (e.g. \citet{Goto03} get
$^{12}$C/$^{13}$C$=86-137$).  One method that is not sensitive to
processes of selective fractionation (described below) is determining
the $^{12}$CH$^+$/$^{13}$CH$^+$ ratio and the recent measurements of
\citet{Casassus05} agree with a value of $\sim$78 but they interpret
the scatter in their data ($1\sigma=\pm12.7$) as a true measure of chemical
heterogeneity in the local ISM.

There are two main processes that can cause $^{12}$CO/$^{13}$CO to
deviate from the average ISM value of the $^{12}$C/$^{13}$C isotope
ratio.  The first is isotopic charge exchange \citep{Watson76}, which
occurs in gas where C$^+$ is in abundance, as might be expected in
translucent clouds, and enhances the $^{13}$CO because of its lower
zero-point energy.  The other process is selective isotopic
photodissociation \citep{Bally82}, which favors the more abundant, and
thus more likely to self-shield, $^{12}$CO.

If isotopic exchange is more important than photodissociation we would
expect $^{12}$CO/$^{13}$CO =
exp($-\Delta E /kT_\mathrm{kin}$)$\times$($^{12}$C/$^{13}$C), where
$\Delta E$ is the zero point energy difference between the two isotopes
($\Delta E/k=35$~K).  The average kinetic temperature for these six
sightlines is 58 K, and therefore we could expect to see as low as
half the average ISM isotopic ratio.  However, it is unlikely that
photodissociation does not play a role, and the models of
\citet{vanDishoeck88} suggest at best only a mild relative increase in
the $^{13}$CO abundance for temperatures in the range seen in this
sample.

\subsection{Rotational Excitation}

Both the \textit{FUSE} and the STIS data are of high enough resolution
to resolve the individual rotational transitions of \htwo\ and CO
respectively.  This allows for the determination of the rotational
excitation temperature.  For both molecules the relative column
densities in the $J=0$ and $J=1$ states are representative of the
rotational excitation temperature of the gas, $T_{01}$, following:

\[N(1)/N(0) = g_1/g_0 \exp(-E_{01}/kT_{01})\]

\noindent where $N(J)$ is the column density in the $J$th rotational
state, $g_J$ is the statistical weight, $E_{01}$ is the energy
difference between the $J=0$ and $J=1$ states, and $k$ is Boltzmann's
constant.

\begin{figure}[t]
\epsscale{1.2}
\plotone{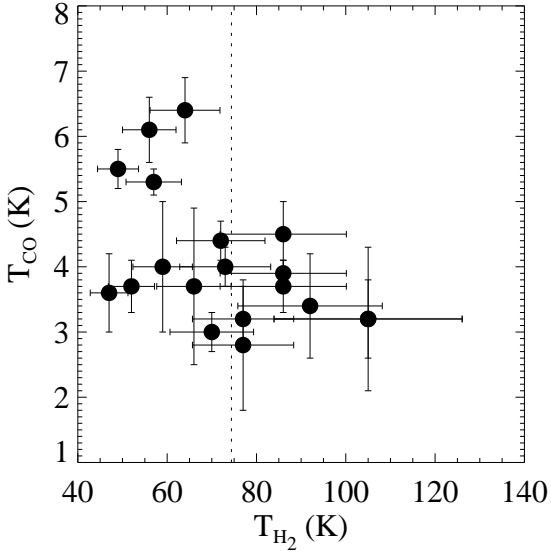}
\caption{CO rotational excitation temperature versus \htwo\ rotational
excitation temperature.  The vertical dotted line is at the average
value for the \htwo\ temperature.  Only sightlines with below average
\htwo\ temperature exhibit enhanced CO temperature.
\label{TCOvTH2}}
\end{figure}

Transitions between the ground-state rotational states of \htwo\ are
not dipole-allowed, and have lifetimes long enough that at the
densities typical of diffuse and translucent clouds collisions
dominate over other processes in determining the relative populations
of the low-lying $J$ levels.  Thus, $T_{01}$ for \htwo\ traces the
kinetic temperature of the gas.  In CO, on the other hand, the
rotational transitions are dipole-allowed and have short lifetimes
such that the rotational excitation temperature will reflect mostly
the density in the cloud, not the kinetic temperature.

The average \htwo\ temperature in our sample is $74\pm24$ K.  This
agrees very well with the value derived from \textit{Copernicus}
observations of 61 stars with \htwo\ columns greater than $10^{18}$
cm$^{-2}$ of $<T_{01}>=77\pm17$ K \citep{Savage77} as well as the
$<T_{01}>=68\pm15$ K of \citet{Rachford02}.  The higher $J$ levels can
be populated through collisions, UV- and formation-pumping, and
radiative cascade and thus their relative populations may not reflect
the kinetic temperature.  The absorptions from these levels will not
be considered in this study and should not greatly affect the column
density determinations because the $J=0$ and 1 absorption accounts for
the vast majority of the \htwo\ column.

There is no correlation of the rotational temperatures of either \htwo\
or CO with reddening along the line of sight, nor was any seen in
either the \textit{Copernicus} study or the \textit{FUSE} study of
Rachford et al.  The rotational temperatures are also independent of
total \htwo\ column density.  However, there are several
interesting relationships with rotational temperature.

\citet{Rachford02} point out that for lines of sight
dominated by translucent clouds, the molecular fraction should be
large while the kinetic temperature is small.  Though we do not see a
strong correlation (i.e. we see several sightlines with low
temperature and molecular fraction) all of the sightlines with above
average \htwo\ temperature have a molecular fraction below 0.2, except
for HD~185418, which has a molecular fraction of 0.49.

For CO column densities less than 10$^{15}$ cm$^{-2}$, the average
rotational excitation temperature is $3.6\pm0.5$ K.  For the higher
column density sightlines, the average is $5.2\pm1.0$ K.  Also, the
average CO rotational temperature is $4.5\pm1.1$ K for sightlines with
below average \htwo\ temperature, and $3.5\pm0.5$ K for those above
(see Figure \ref{TCOvTH2}).  The increase of excitation with column
density may be a result of ``photon trapping'' in clouds of higher
optical depth.  As the optical depth of the 2.6~mm $J=1-0$ radio line
increases, it becomes more likely that the photon will be reabsorbed
by another CO molecule.  This hinders the radiative cooling and the
collisional excitation/de-excitation process with the surrounding
\htwo\ molecules will equilibrate at a higher rotational temperature.
The CO $T_\mathrm{rot}$ is thus sensitive to the \htwo\ space density,
the \htwo\ kinetic temperature, and the line center optical depth,
which itself depends on the column density and $b$-value.

Using an on-line statistical equilibrium radiative transfer code,
available at\\
 \url{http://www.strw.leidenuniv.nl/$\sim$moldata}
\citep{Schoier05}, we determined that the increase in temperature for
the higher column density sightlines is consistent with the ``photon
trapping'' effect in clouds with \htwo\ space densities in the
20-100~cm$^{-3}$ range, assuming our average \htwo\ temperature as the
kinetic temperature and a $b$-value of about 0.6~km~s$^{-1}$.  This
code assumes a single, homogeneous cloud, which may not necessarily
represent a sightline with multiple velocity components.

Another method for increasing the rotational excitation of
CO is that of radiative pumping from nearby dense clouds, as
studied by \citet{Wannier97}.  This effect will be proportional to the
solid angle subtended by the nearby cloud as viewed from the CO along
the line of sight, when the relative velocity between the absorbing
and emitting clouds is zero.  They point out that for sightlines with
space densities in the few hundreds per cm$^{3}$, and with kinetic
temperatures in the range of the \htwo\ temperatures we observe,
fractions of $\sim$0.2 of 4$\pi$ steradians is enough to account
for the increased CO temperatures observed.

\citet{Wannier97} point out that this effect may be differentiated
from collisional excitation by observing other isotopes of CO, because
the lower brightness temperature of the nearby clouds will reduce the
rate of radiative pumping.  Table 3 lists the $^{13}$CO temperatures
derived in this study, with the isotopic temperature ratio
$T_{12}/T_{13}$ listed in the final column.  For two or possibly three
of the sightlines, there is an increased relative temperature for the
more abundant isotope.  These are also the sightlines with the higher
molecular fractions, and are perhaps sightlines that pass through the
outer portions of a denser molecular cloud, where a larger fraction of
its sky may be covered by optically thick, radiating gas.  However, we
note that the ``photon trapping'' could also explain the
$T_{12}/T_{13}$, because the less optically-thick $^{13}$CO would not
have its rotational temperature enhanced by the effect.

\section{Discussion}

The traditional view of the ISM differentiates between diffuse and
translucent clouds, where the transition occurs at about $A_V=1$.
However, recent studies \citep{Rachford02,Sonnentrucker03} have
suggested that there is not a clear distinction, and argue that
perhaps none of the sightlines they see have truly ``translucent''
clouds, which models expect to show molecular fractions reaching very
nearly 1.  This situation could arise from having multiple ``diffuse''
clouds along the line of sight, adding up to a given $A_V$.  This
would result in \ebv\ not necessarily being a good measure of the
physical properties along a given line of sight, and our study
confirms this.

It is thus more important to measure parameters that are better
representative of the physical state of the clouds.  The molecular
fraction gives one such measure, but for reasons laid out in
\citet{Rachford02} may not allow for the unambiguous separation of
diffuse and translucent sightlines.  However, the CO molecule is more
sensitive to photodissociative processes, and can be used to probe the
transition region between diffuse and translucent clouds.
\citet{Kopp00} model the sensitivity of the abundance of CO to the
effects of geometry, dust shielding and fragmentation of the
ISM.  Their results suggest that if a sightline were
simply a collection of small, diffuse clouds we would not expect to
see an increased CO/\htwo\ with total column density.  The
increase in CO/\htwo\ that we observe with increased
molecular fraction suggests that these sightlines do begin to sample
the translucent regime.

We believe that the transition from low to high CO/\htwo\ is similar
to the one that hydrogen undergoes when the \htwo\ column density gets
high enough for self-shielding to take effect.  The molecular fraction
observed in a cloud will be dependent on the balance of the formation
and the photodestruction.  \citet{Savage77} found that the molecular
fraction transitions from ``low'' ($<0.01$) to ``high'' ($>0.01$)
values at about $N_H=5\times10^{20}$~cm$^{-2}$, which corresponds to
$E(B-V)\sim0.08$.  This is also seen for high-latitude Galactic
sightlines \citep{Gillmon06}.  The location of this transition is
dependent on the physical conditions in the ISM, including the \htwo\
formation rate on grains, the intensity of the interstellar radiation
field, and the degree of fragmentation of molecular clouds along the
line of sight.  The behavior of the Galactic sightlines is in marked
contrast to that of the Magellanic Clouds \citep{Tumlinson02}, in
which this transition occurs at much larger columns, perhaps because
of the lower \htwo\ formation rates and higher dissociating radiation
fields.  All of the sightlines in this study have total hydrogen
columns above this transition value, and should be considered part of
the ``high'' molecular fraction regime.

The values of CO/\htwo\ that we measure for the diffuse and
translucent regimes range from $10^{-7}-10^{-5}$, in contrast with the
canonical dense cloud value of $10^{-4}$.  This discrepancy is
consistent with the studies of \citet{deVries87} and
\citet{Magnani98}, which find variations in the X-factor as determined
from radio observations of high-latitude translucent clouds.  Both
studies interpret their results as variations of the CO abundance,
which could have a deleterious effect on the process of determining
the masses of individual clouds.  It could also impact the "weighing"
of galaxies if the diffuse or translucent ISM contribute significantly
to the CO radio emission.  \citet{Polk88} found, based on the ratios
of integrated $^{12}$CO and $^{13}$CO emission, that a significant
contribution to the total $^{12}$CO emission from a galaxy could be
arising from material of moderate optical depth.  In that case,
applying a standard X-factor, derived from the dense clouds, would
result in the underestimation of the total mass because of the lower
CO/\htwo\ in the more diffuse gas.

This difference could be particularly important for clouds subjected to
intense UV radiation fields.  \citet{Yao03} found in their study of CO 
emission in starburst galaxies an X-factor significantly lower than
the standard Galactic value.  Further, they found, by virial analysis,
that the clouds in these galaxies were not gravitationally bound unless
the CO/\htwo\ was 9-90 times lower than expected and suggest that the
CO emission arises from nonvirialized warm and diffuse gas clouds.  On
the other hand, \citet{Rosolowsky03} find no evidence in M33 for a diffuse
molecular component traced by CO emission, concluding that most of the CO
flux resides in the giant molecular clouds.

The direct comparison of the column densities determined from UV
absorption line spectroscopy of CO and \htwo\ provides a technique for
probing the structure and chemical balance of the ISM free from many
of the biases and assumptions inherent in the traditional X-factor
method.  If the CO J=1-0 radio emission along these lines of sight
could be measured then the relationship between CO/\htwo\ and X-factor
could be checked directly.  It would also be useful to use absorption
spectroscopy to explore the sightlines with
$N($\htwo)$=10^{21}-10^{22}$~cm$^{-2}$, the regime in which we expect
to see the transition from translucent to dense clouds.  The power law
increase in CO versus \htwo\ seen in the diffuse/translucent regime
should turn over as $A_V$ rises above $\sim3$ and the CO/\htwo\
``saturates'' at the canonical dense cloud value of $\sim10^{-4}$.

Furthermore, extragalactic environments could be explored, such as the
Magellanic Clouds; however, we do expect that with the currently
available archival data CO will be difficult to detect there.  The
study of \citet{Tumlinson02} found that although \htwo\ column
densities as high as around $10^{20}$~cm$^{-2}$ have been detected, the
typical columns found in the Clouds lie in the
$10^{15}-10^{19}$~cm$^{-2}$ range.  If the CO/\htwo\ ratio were to
follow the trend seen in the Galactic sample, the CO column densities
would be below the STIS detection threshold.  We would expect, though,
that the CO/\htwo\ would be even lower because of the lower
metallicity and higher UV radiation fields seen in the Clouds.
But, because models predict \citep[e.g.]{Bell06} the CO/\htwo\ to change
with environment, if the appropriate data were taken, the technique
could be used to explore the varying conditions in the
diffuse/translucent ISM in other galaxies as well as our own.

\begin{acknowledgements}
The authors would like to thank Chris Howk for helping with the
initial archive search and B-G Andersson for valuable input.  EBB
would like to thank Jay Gallagher and John Mathis for useful
discussions.  The authors would also like to thank the referee for
very useful suggetions, particularly those that led to a more robust
fitting method used in fitting high column density sightlines with
unresolved velocity structure.  All of the data presented in this
paper were obtained from the Multimission Archive at the Space
Telescope Science Institute (MAST). STScI is operated by the
Association of Universities for Research in Astronomy, Inc., under
NASA contract NAS5-26555. Support for MAST for non-HST data is
provided by the NASA Office of Space Science via grant NAG5-7584 and
by other grants and contracts.
\end{acknowledgements}

\input{tab1.tex}
\input{tab2.tex}
\input{tab3.tex}

\end{document}

%% file: tab1.tex
\begin{deluxetable}{lcc|ccc|cc|r@{\extracolsep{0pt}.}l}
\small
\tablewidth{7.0in}
\tablecaption{Molecular Column Densities and Rotational Temperatures}
\tablehead{&&&\multicolumn{3}{c|}{CO}&\multicolumn{2}{c|}{H$_2$}&\multicolumn{2}{c}{
}\\
\cline{4-8}
 & & &log $N$& $T_{\mathrm{rot}}$
& $b$  & log $N^a$ & $T_{\mathrm{rot}}$&\multicolumn{2}{c}{CO/H$_2$}\\
Star & Sp. Type & $E_{B-V}$ & (cm$^{-2}$) & (K) & (km~s$^{-1}$) &(cm$^{-2}$)&
(K)& \multicolumn{2}{c}{($\times10^{-6}$)} }
\startdata
HD 24534$^b$ & B0Ve & 0.45 & 16.13$\pm$0.20 & 5.3$\pm$0.6 &\phn0.4$\pm$0.2     & 20.92&\phn57$\pm$6\phn&16&1\phn$\pm$8.3\\
HD 27778$^b$  & B3V & 0.38 & 16.05$\pm$0.14 & 6.1$\pm$0.5 & 0.7$_{-0.1}^{+0.3}$& 20.79&\phn56$\pm$6\phn&18&0\phn$\pm$7.1\\
HD 91824 & O7V     & 0.27 & $\lesssim$ 13.60& -           & -                  & 19.99&\phn61$\pm$7\phn&$\lesssim0$&40\\
HD 93205 & O3V      & 0.37 & 13.23$\pm$0.06 & 3.4$\pm$0.6 &\phn4.3$\pm$1.0     & 19.84&   105$\pm$21   &0&24$\pm$0.07\\
HD 93222 & O7III    & 0.40 & 13.36$\pm$0.20 & 2.8$\pm$1.0 &\phn0.7$\pm$0.6     & 19.80&\phn77$\pm$11   &0&36$\pm$0.19\\
HD 93840 & B1.0Ib   & 0.14 & 13.38$\pm$0.08 & 3.6$\pm$0.6 & 1.1$_{-0.5}^{+0.8}$& 19.55&\phn47$\pm$4\phn&0&68$\pm$0.21\\
HD 102065$^c$ & B2V & 0.31 & 13.62$\pm$0.12 & 4.0$\pm$1.0 &\phn1.7$\pm1.2$     & 20.63&\phn59$\pm$7\phn&0&10$\pm$0.04\\
HD 103779 &B0.5III  & 0.21 & $\lesssim$12.35& -           & -                  & 19.90&\phn86$\pm$14   &$\lesssim0$&03\\
HD 104705 & B0Ib    & 0.26 & 12.98$\pm$0.16 & 3.4$\pm$0.8 & 1.0$_{-0.5}^{+1.2}$& 19.98&\phn92$\pm$16   &0&10$\pm$0.04\\
HD 116852 & O9IV    & 0.22 & 13.28$\pm$0.04 & 3.0$\pm$0.3 & 0.4$_{-0.1}^{+0.5}$& 19.85&\phn70$\pm$9\phn&0&27$\pm$0.07\\
HD 121968 & B1V     & 0.07 & $\lesssim$12.30& -           & -                  & 18.70&\phn38$\pm$3\phn&$\lesssim 0$&40\\
HD 152723 &O6.5III  & 0.46 & 13.88$\pm$0.15 & 4.0$\pm$0.3 &\phn1.0$\pm$0.2     & 20.33&\phn73$\pm$10   &0&36$\pm$0.15\\
HD 163758 &O6.5Iaf  & 0.35 & 13.42$\pm$0.05 & 4.5$\pm$0.5 &\phn1.3$\pm$0.5     & 19.85&\phn86$\pm$14   &0&37$\pm$0.10\\
HD 177989 &B0III    & 0.25 & 14.62$\pm$0.17 & 3.7$\pm$0.4 &\phn0.7$\pm$0.1     & 20.23&\phn52$\pm$5\phn&2&4\phn$\pm$1.1\\
HD 185418 &B0.5V    & 0.50 & 14.82$\pm$0.20 & 3.2$\pm$1.1 & 0.8$_{-0.3}^{+1.0}$& 20.79&   105$\pm$21   &1&1\phn$\pm$0.6\\
HD 201345 & O9p     & 0.32 & $\lesssim$12.40& -           & -                  & 19.43&   147$\pm$41   &$\lesssim0$&09\\
HD 203532 &B3IV     & 0.32 & 15.70$\pm$0.17 & 5.5$\pm$0.3 &\phn0.7$\pm$0.1     & 20.71&\phn49$\pm$5\phn&9&8\phn$\pm$4.5\\
HD 206267$^b$&O6.5V & 0.52 & 16.11$\pm$0.13 & 6.4$\pm$0.5 & 0.7$_{-0.1}^{+0.8}$& 20.86&\phn64$\pm$8\phn&17&9\phn$\pm$6.8\\
HD 207198$^b$&O9.5Ib& 0.62 & 15.53$\pm$0.20 & 3.7$\pm$1.2 &\phn1.0$\pm$0.5     & 20.83&\phn66$\pm$8\phn&5&0\phn$\pm$2.6\\
HD 210839$^b$&O6Infp& 0.56 & 15.41$\pm$0.08 & 4.4$\pm$0.3 & 0.9$_{-0.1}^{+0.2}$& 20.84&\phn72$\pm$10   &3&7\phn$\pm$1.1\\
HD 218915 &O9.5Iab  & 0.29 & 13.64$\pm$0.13 & 3.9$\pm$0.2 &\phn1.6$\pm$0.6     & 20.15&\phn86$\pm$14   &0&31$\pm$0.12\\	
HD 303308 & B1.0III & 0.30 & 13.65$\pm$0.06 & 3.7$\pm$0.4 &\phn3.1$\pm$0.5     & 20.35&\phn86$\pm$14   &0&20$\pm$0.05\\
CPD -59 2603 &O7V   & 0.46 & 14.15$\pm$0.10 & 3.2$\pm$0.5 & 0.6$_{-0.1}^{+0.3}$& 20.15&\phn77$\pm$11   &1&00$\pm$0.33\\
\enddata
\tablenotetext{a}{Errors are $\pm0.10$}
\tablenotetext{b}{\htwo\ values from Rachford et al. (2002)}
\tablenotetext{c}{Sp. Type and $E(B-V)$ determined this study}
\end{deluxetable}

%% file: tab2.tex
\begin{deluxetable}{lcc|cc|c}
\small
\tablewidth{6in}
\tablecaption{Molecular and Atomic Hydrogen Column Densities}
\tablehead{& \multicolumn{2}{c}{H$_2$ Column Densities (cm$^{-2}$)$^{a}$} & \multicolumn{2}{c}{\ion{H}{1}} & Molecular\\
Star & log $N$(0) & log $N$(1) & log $N$ (cm$^{-2}$) & Ref. & Fraction}
\startdata
HD 24534$^b$& 20.76 & 20.42 & 20.73 &1&0.76\\
HD 27778$^b$& 20.64 & 20.27 & 21.20 &4&0.44\\
HD 91824    & 19.80 & 19.55 & 21.15 &2&0.12\\
HD 93205    & 19.40 & 19.65 & 21.33 &1&0.06\\
HD 93222    & 19.50 & 19.50 & 21.54 &2&0.04\\
HD 93840    & 19.45 & 18.85 & 21.04 &1&0.06\\
HD 102065   & 20.45 & 20.15 & 21.90 &4&0.10\\
HD 103779   & 19.55 & 19.65 & 21.16 &1&0.10\\
HD 104705   & 19.60 & 19.75 & 21.11 &1&0.13\\
HD 116852   & 19.60 & 19.50 & 20.96 &1&0.14\\
HD 121968   & 18.65 & 17.70 & 20.71 &1&0.02\\
HD 152723   & 20.05 & 20.00 & 21.43 &1&0.14\\
HD 163758   & 19.50 & 19.60 & 21.23 &1&0.08\\
HD 177989   & 20.10 & 19.65 & 20.95 &1&0.28\\
HD 185418   & 20.35 & 20.60 & 21.11 &2&0.49\\
HD 201345   & 18.85 & 19.30 & 20.88 &1&0.07\\
HD 203532   & 20.60 & 20.05 & 21.50 &4&0.24\\
HD 206267$^b$& 20.64& 20.45 & 21.30 &3&0.42\\
HD 207198$^b$& 20.61& 20.44 & 21.34 &1&0.38\\
HD 210839$^b$& 20.57& 20.50 & 21.15 &1&0.53\\
HD 218915   & 19.80 & 19.90 & 21.11 &1&0.18\\
HD 303308   & 20.00 & 20.10 & 21.45 &1&0.14\\
CPD -59 2603& 19.85 & 19.85 & 21.46 &1&0.09\\
\enddata
\tablenotetext{a}{Uncertainties for N(0,1) $\approx \pm0.1$}
\tablenotetext{b}{\htwo\ values from Rachford et al. (2002)}
\tablerefs{(1) \citet{Diplas94}, (2) \citet{FM90}, (3) \citet{Rachford02}, (4) Determined this study}
\end{deluxetable}

%% file: tab3.tex
\begin{deluxetable}{l|cc|c|cc|cc}
\small
\tablewidth{6.2in}
\tablecaption{CO Isotopic Column Densities and Rotational Temperatures}
\tablehead{&\multicolumn{2}{|c|}{$^{12}$CO}&\multicolumn{3}{c|}{$^{13}$CO}&&\\
\cline{2-6}
&Log $N$&$T_{\mathrm{rot}}$&Log $N$$^a$&Log $N$&$T_{\mathrm{rot}}$&&\\
Star&(cm$^{-2}$)&(K)&(cm$^{-2}$)&(cm$^{-2}$)&(K)&$^{12}$CO/$^{13}$CO&$T_{12}/T_{13}$}
\startdata
HD 24534 &16.13$\pm$0.20&5.3$\pm$0.6&14.33&14.30$\pm$0.12&4.1$\pm$0.6&68$\pm$31 & 1.3$\pm$0.3\\
HD 27778 &16.05$\pm$0.13&6.1$\pm$0.5&14.26&14.28$\pm$0.08&3.9$\pm$0.6&59$\pm$14 & 1.6$\pm$0.3\\
HD 177989&14.62$\pm$0.17&3.7$\pm$0.2&12.90&12.82$\pm$0.08&3.9$\pm$0.5&63$\pm$25 & 1.0$\pm$0.2\\
HD 203532&15.70$\pm$0.17&5.5$\pm$0.5&13.95&13.97$\pm$0.20&4.8$\pm$0.6&54$\pm$21 & 1.1$\pm$0.2\\
HD 206267&16.11$\pm$0.17&6.4$\pm$0.6&14.44&14.42$\pm$0.08&6.4$\pm$0.8&49$\pm$15 & 1.0$\pm$0.2\\
HD 210839&15.41$\pm$0.04&4.5$\pm$0.2&13.75&13.70$\pm$0.10&3.5$\pm$0.8&51$\pm$9\phn & 1.3$\pm$0.3\\
\enddata
\tablenotetext{a}{Column density assuming same $T_\mathrm{rot}$ as for $^{12}$CO}
\end{deluxetable}

%% file: emapj.bbl
\begin{thebibliography}{}

\bibitem[Bally \& Langer(1982)]{Bally82} Bally, J., \& Langer, 
W.~D.\ 1982, \apj, 255, 143 

\bibitem[Bell et al.(2006)]{Bell06} Bell, T.~A., Roueff, E., 
Viti, S., \& Williams, D.~A.\ 2006, ArXiv Astrophysics e-prints, 
arXiv:astro-ph/0607428 

\bibitem[Bohlin, Savage \& Drake(1978)]{Bohlin78} Bohlin, R.~C., Savage, 
B.~D., \& Drake, J.~F.\ 1978, \apj, 224, 132 
 
\bibitem[Burgh et al.(2000)]{Burgh00} Burgh, E.~B., McCandliss, S.~R., 
Andersson, B.-G., \& Feldman, P.~D.\ 2000, \apj, 541, 250 

\bibitem[Casassus et al.(2005)]{Casassus05} Casassus, S., Stahl, 
O., \& Wilson, T.~L.\ 2005, \aap, 441, 181 

\bibitem[Crenny \& Federman(2004)]{Crenny04} Crenny, T., \& 
Federman, S.~R.\ 2004, \apj, 605, 278 

\bibitem[de Vries et al.(1987)]{deVries87} de Vries, H.~W., 
Thaddeus, P., \& Heithausen, A.\ 1987, \apj, 319, 723 

\bibitem[Dickman(1978)]{Dickman78} Dickman, R.~L.\ 1978, \apjs, 
37, 407 

\bibitem[Diplas \& Savage(1994)]{Diplas94} Diplas, A., \& 
Savage, B.~D.\ 1994, \apjs, 93, 211 

\bibitem[Federman et al.(1980)]{Federman80} 
Federman, S.~R., Glassgold, A.~E., Jenkins, E.~B., \& Shaya, E.~J.\ 1980, 
\apj, 242, 545 

\bibitem[Federman et al.(2003)]{Federman03} Federman, S.~R., Lambert,
D.~L., Sheffer, Y., Cardelli, J.~A., Andersson, B.-G., van Dishoeck,
E.~F., \& Zsarg{\'o}, J.\ 2003, \apj, 591, 986

\bibitem[Fitzpatrick \& Massa(1990)]{FM90} Fitzpatrick, 
E.~L., \& Massa, D.\ 1990, \apjs, 72, 163 

\bibitem[Gillmon et al.(2006)]{Gillmon06} Gillmon, K., Shull, 
J.~M., Tumlinson, J., \& Danforth, C.\ 2006, \apj, 636, 891 

\bibitem[Goto et al.(2003)]{Goto03} Goto, M., et al.\ 2003, 
\apj, 598, 1038 

\bibitem[Jenniskens \& Greenberg(1993)]{Jenniskens93} Jenniskens, 
P., \& Greenberg, J.~M.\ 1993, \aap, 274, 439 

\bibitem[Kaczmarczyk(2000)]{Kaczmarczyk00} Kaczmarczyk, G.\ 2000, 
\mnras, 316, 875 

\bibitem[Kopp, Roueff, \& Pineau des For{\^ 
e}ts(2000)]{Kopp00} Kopp, M., Roueff, E., \& Pineau des For{\^ 
e}ts, G.\ 2000, \mnras, 315, 37 

\bibitem[Lacy et al.(1994)]{Lacy94} Lacy, J.~H., Knacke, R., Geballe T.~R., \& Tokunaga, A.~T.\ 1994, \apjl, 428, L69

\bibitem[Lambert et al.(1994)]{Lambert94} Lambert, D.~L., 
Sheffer, Y., Gilliland, R.~L., \& Federman, S.~R.\ 1994, \apj, 420, 756 

\bibitem[Langer \& Penzias(1993)]{Langer93} Langer, W.~D., \& 
Penzias, A.~A.\ 1993, \apj, 408, 539 

\bibitem[Magnani et al.(1998)]{Magnani98} Magnani, L., Onello, 
J.~S., Adams, N.~G., Hartmann, D., \& Thaddeus, P.\ 1998, \apj, 504, 290 

\bibitem[McCandliss(2003)]{McCandliss03} McCandliss, S.~R.\ 2003, 
\pasp, 115, 651 

\bibitem[Morton \& Noreau(1994)]{Morton94} Morton, D.~C.~\& 
Noreau, L.\ 1994, \apjs, 95, 301 

\bibitem[Pan et al.(2004)]{Pan04} Pan, K., Federman, S.~R., 
Cunha, K., Smith, V.~V., \& Welty, D.~E.\ 2004, \apjs, 151, 313 

\bibitem[Pan et al.(2005)]{Pan05} Pan, K., Federman, S.~R., 
Sheffer, Y., \& Andersson, B.-G.\ 2005, \apj, 633, 986

\bibitem[Polk et al.(1988)]{Polk88} Polk, K.~S., Knapp, G.~R., 
Stark, A.~A., \& Wilson, R.~W.\ 1988, \apj, 332, 432 

\bibitem[Rachford et al.(2002)]{Rachford02} Rachford, B.~L.~et 
al.\ 2002, \apj, 577, 221 

\bibitem[Rosolowsky et al.(2003)]{Rosolowsky03} Rosolowsky, E., 
Engargiola, G., Plambeck, R., \& Blitz, L.\ 2003, \apj, 599, 258 

\bibitem[Savage et al.(1977)]{Savage77} 
Savage, B.~D., Bohlin, R.~C., Drake, J.~F., \& Budich, W.\ 1977, \apj, 216, 
291 

\bibitem[Sch{\"o}ier et al.(2005)]{Schoier05} Sch{\"o}ier, F.~L., 
van der Tak, F.~F.~S., van Dishoeck, E.~F., \& Black, J.~H.\ 2005, \aap, 
432, 369 

\bibitem[Sheffer, Federman, \& Lambert(2002)]{Sheffer02} Sheffer, Y., Federman, S.~R., \& Lambert, D.~L.\ 2002,\apjl, 572, L95

\bibitem[Sonnentrucker et al.(2003)]{Sonnentrucker03} Sonnentrucker, 
P., Friedman, S.~D., Welty, D.~E., York, D.~G., \& Snow, T.~P.\ 2003, \apj, 
596, 350 

\bibitem[Shull \& Beckwith(1982)]{Shull82} Shull, J.~M.~\& 
Beckwith, S.\ 1982, \araa, 20, 163 

\bibitem[Spitzer et al.(1974)]{Spitzer74} Spitzer, L., Cochran, 
W.~D., \& Hirshfeld, A.\ 1974, \apjs, 28, 373 

\bibitem[Strong \& Mattox(1996)]{Strong96} Strong, A.~W., \& 
Mattox, J.~R.\ 1996, \aap, 308, L21 

\bibitem[Tumlinson et al.(2002)]{Tumlinson02} Tumlinson, J., et 
al.\ 2002, \apj, 566, 857 

\bibitem[Valencic et al.(2004)]{Valencic04} Valencic, L.~A., 
Clayton, G.~C., \& Gordon, K.~D.\ 2004, \apj, 616, 912 

\bibitem[van Dishoeck \& Black(1988)]{vanDishoeck88} van Dishoeck, 
E.~F.~\& Black, J.~H.\ 1988, \apj, 334, 771 

\bibitem[Walborn et al.(1995)]{Walborn95} Walborn, N.~R., Parker, 
J.~W., \& Nichols, J.~S.\ 1995, $IUE$ Atlas of B-type Spectra from 1200 to 1900 \AA\ (NASA RP-1363)

\bibitem[Wannier, Penprase, \& Andersson(1997)]{Wannier97} 
Wannier, P., Penprase, B.~E., \& Andersson, B.-G.\ 1997, \apjl, 487, L165 

\bibitem[Watson et al.(1976)]{Watson76} Watson, W.~D., Anicich, 
V.~G., \& Huntress, W.~T.\ 1976, \apjl, 205, L165 
 
\bibitem[Wilson \& Rood(1994)]{Wilson94} Wilson, T.~L., \& Rood, R.\ 1994, \araa, 32, 191

\bibitem[Yao et al.(2003)]{Yao03} Yao, L., Seaquist, E.~R., 
Kuno, N., \& Dunne, L.\ 2003, \apj, 588, 771 

\bibitem[Young \& Scoville(1991)]{Young91} Young, J.~S., \& 
Scoville, N.~Z.\ 1991, \araa, 29, 581 

\end{thebibliography}
